\documentclass[sigconf]{acmart}

\setcopyright{acmlicensed}
\copyrightyear{2026}
\acmConference[ICONS '26]{International Conference on Neuromorphic Systems}{2026}{Chicago, IL}

\usepackage{amsmath}
\usepackage{algorithmic}
\usepackage{graphicx}
\usepackage{textcomp}
\usepackage{xcolor}
\usepackage{booktabs}
\usepackage{url}

\usepackage{tikz}
\usetikzlibrary{arrows.meta, decorations.pathreplacing}

\begin{document}

\title{FlowEqProp: Training Flow Matching Generative Models with Gradient Equilibrium Propagation}

\author{Alex Gower}
\authornote{This work was supported by UKRI/EPSRC CASE Award under Grant 220191 in partnership with Nokia UK Limited.}
\email{apg59@cam.ac.uk}
\affiliation{%
  \institution{Nokia Bell Labs \& University of Cambridge}
  \city{Cambridge}
  \country{United Kingdom}
}



\begin{abstract}
We introduce Gradient Equilibrium Propagation (GradEP), a mechanism that extends Equilibrium Propagation (EP) to train energy gradients rather than energy minima, enabling EP to be applied to tasks where the learning objective depends on the velocity field of a convergent dynamical system. Instead of fixing the input during dynamics as in standard EP, GradEP introduces a spring potential that allows all units, including the visible units, to evolve, encoding the learned velocity in the equilibrium displacement. The spring and resulting nudge terms are both purely quadratic, preserving EP's hardware plausibility for neuromorphic implementation. As a first demonstration, we apply GradEP to flow matching for generative modelling --- an approach we call FlowEqProp --- training a two-hidden-layer MLP (24,896 parameters) on the Optical Recognition of Handwritten Digits dataset using only local equilibrium measurements and no backpropagation. The model generates recognisable digit samples across all ten classes with stable training dynamics. We further show that the time-independent energy landscape enables extended generation beyond the training horizon, producing sharper samples through additional inference-time computation --- a property that maps naturally onto neuromorphic hardware, where longer relaxation yields higher-quality outputs. To our knowledge, this is the first demonstration of EP training a flow-based generative model.
\end{abstract}


\maketitle

\section{Introduction}\label{sec:introduction}

Neuromorphic hardware --- including analog circuits, memristive crossbar arrays, and oscillator-based processors --- promises orders-of-magnitude improvements in speed and energy efficiency for neural computation by exploiting physical dynamics directly rather than simulating them digitally~\cite{gowerHowTrainOscillator_Published}. However, realising this potential requires training algorithms compatible with the constraints of physical systems: local computation and no separate backward pass. Equilibrium Propagation (EP)~\cite{scellierEquilibriumPropagationBridging2017} satisfies these requirements for convergent dynamical systems, extracting parameter gradients from local measurements at equilibrium, making it one of the most promising candidates for on-chip learning.

To date, EP has been demonstrated primarily on classification tasks~\cite{laborieuxScalingEquilibriumPropagation2020, ernoultUpdatesEquilibriumProp2019},\footnote{A recent extension applies EP to variational autoencoders~\cite{meerschTrainingHopfieldVariational2023}.} where inputs are clamped and the loss acts on the settled output. Flow-based generative modelling --- where the training objective acts on the \textit{velocity} (energy gradient) of the dynamical system rather than on settled state values --- has remained out of reach, as standard EP provides no mechanism for training energy gradients. In particular, flow matching~\cite{lipmanFlowMatchingGenerative2023}, which generalises diffusion-based approaches and achieves state-of-the-art generation quality, currently relies on backpropagation for training.

In this work, we introduce \textit{Gradient Equilibrium Propagation} (GradEP), a general mechanism for training energy gradients via EP. By replacing the hard input clamp of standard EP with a spring potential, GradEP allows all units --- including visible units --- to evolve during dynamics, so that the equilibrium displacement directly encodes the learned velocity. The resulting nudge loss on the velocity is purely quadratic and requires no backpropagation through the energy function, maintaining compatibility with neuromorphic hardware. As a first demonstration, we apply GradEP to flow matching for generative modelling --- an approach we call \textit{FlowEqProp} --- and generate recognisable digit samples across all ten classes using a two-hidden-layer MLP with only 24,896 parameters on the Optical Recognition of Handwritten Digits dataset. Moreover, the time-independent energy landscape enables extended generation beyond $t = 1$, producing sharper samples through additional inference-time computation. To our knowledge, this is the first demonstration of EP training a flow-based generative model. Our contributions are:

\begin{enumerate}
    \item \textbf{GradEP}: a general mechanism for training energy gradients via EP, applicable to flow matching~\cite{lipmanFlowMatchingGenerative2023}, score matching~\cite{songGenerativeModelingEstimating2019}, and energy-based generation~\cite{balcerakEnergyMatchingUnifying2025}.
    \item \textbf{FlowEqProp}: the first EP-trained flow matching model, producing recognisable digit samples with stable training dynamics.
\end{enumerate}

\section{Background}\label{sec:background}

\subsection{Equilibrium Propagation}\label{sec:ep}

Equilibrium Propagation (EP)~\cite{scellierEquilibriumPropagationBridging2017,laborieuxScalingEquilibriumPropagation2020} is a learning algorithm for convergent recurrent neural networks whose dynamics settle to stationary points. EP computes parameter gradients using only local measurements at equilibrium and requires no separate backward pass (unlike backpropagation), making it naturally suited to neuromorphic hardware. When simulated on conventional hardware, EP also requires only $O(1)$ memory compared to $O(T)$ for BPTT, where $T$ is the number of convergence steps.

The core idea is as follows. Consider a system with dynamical state $s$, trainable parameters $\theta$, and an internal energy $E(s, \theta)$. During inference, the state evolves by gradient descent $\dot{s} = -\partial E / \partial s$, settling to an equilibrium $s_*$ that determines the system's output. Training therefore amounts to shaping $E$ such that its minima encode correct outputs. To achieve this, EP introduces a total energy $F = E + \beta\,\ell$ during training, where $\ell(s)$ is a differentiable loss function and $\beta$ is a scalar nudging factor, and evolves $\dot{s} = -\partial F / \partial s$. We denote $\mathcal{L} = \ell(s_*)$, the loss evaluated at equilibrium. For each training example, EP proceeds in three phases~\cite{laborieuxScalingEquilibriumPropagation2020}: (1) a \textbf{free phase} ($\beta=0$), settling to equilibrium $s_*$; (2) a \textbf{positive nudge} ($\beta>0$), displacing the equilibrium from $s_*$ toward lower loss at $s_*^{+\beta}$; and (3) a \textbf{negative nudge} ($\beta<0$), displacing it toward higher loss at $s_*^{-\beta}$. Parameters are updated via:
\begin{equation}\label{eq:ep-update}
    \Delta \theta \propto -\frac{1}{2\beta} \left( \frac{\partial F}{\partial \theta}\bigg|_{s_*^{+\beta}} - \frac{\partial F}{\partial \theta}\bigg|_{s_*^{-\beta}} \right),
\end{equation}

which decreases energy at $s_*^{+\beta}$ and increases it at $s_*^{-\beta}$, thereby steering the free equilibrium $s_*$ toward lower loss. This recovers exact gradient descent on $\mathcal{L}$ in the limit $\beta \to 0$, $\lim_{\beta \to 0} \Delta\theta \propto -\partial \mathcal{L} / \partial \theta$~\cite{laborieuxScalingEquilibriumPropagation2020}. Each term $\partial F / \partial \theta$ depends only on locally available quantities at connected neurons, so the entire procedure can be implemented without global backward passes.

For example, in classification, $s = (h, y)$ comprises hidden and output units evolving under a fixed input $x$, and $\ell$ measures the discrepancy between the output $y$ and a target label $\hat{y}$. To date, EP has been applied primarily to such classification tasks where inputs are clamped and only hidden/output units evolve~\cite{laborieuxScalingEquilibriumPropagation2020, ernoultUpdatesEquilibriumProp2019}. In this work, we introduce Gradient Equilibrium Propagation (GradEP), extending EP to generative modelling, where all units, including visible units, evolve during dynamics, and the loss acts on the \textit{velocity} of visible units rather than their settled values. This requires a fundamentally different clamping mechanism, described in Section~\ref{sec:method}.

\subsection{Flow Matching}\label{sec:fm}

Flow Matching~\cite{lipmanFlowMatchingGenerative2023} is a framework for training generative models by learning a time-dependent velocity field $v_t(x; \theta)$ that transports noise $x_0 \sim p_0 = \mathcal{N}(0, I)$ at $t=0$ to data $x_1 \sim p_1 \approx q(x_1)$ at $t=1$. It generalises diffusion-based approaches, which correspond to specific (typically curved) choices of transport path.

The true marginal velocity field $u_t(x)$ that generates this transport is intractable. However, a key result of~\cite{lipmanFlowMatchingGenerative2023} is that it can be decomposed into per-sample \textit{conditional} velocity fields $u_t(x \mid x_1)$, and that regressing against these conditional fields yields identical parameter gradients to regressing against the marginal. In practice, training reduces to sampling $t \sim \mathcal{U}[0,1]$, noise $x_0 \sim p_0$, and data $x_1 \sim q(x_1)$, constructing an interpolation $x_t$, and regressing $v_t(x_t; \theta)$ onto the conditional target $u_t(x_t \mid x_1)$.

Using optimal transport (OT) conditional paths, this interpolation is a straight line:
\begin{equation}\label{eq:ot-interp}
    x_t = (1 - t)\, x_0 + t\, x_1,
\end{equation}
with constant conditional velocity $\hat{v} := x_1 - x_0$.
Compared to the curved paths arising from diffusion processes, these straight-line trajectories are simpler to learn and more efficient to sample from~\cite{lipmanFlowMatchingGenerative2023}. The training loss reduces to:
\begin{equation}\label{eq:cfm-ot}
    \mathcal{L}_{\text{CFM}}(\theta) = \mathbb{E}_{t \sim \mathcal{U}[0,1],\, x_0,\, x_1} \left\| v_t(x_t; \theta) - (x_1 - x_0) \right\|^2.
\end{equation}
After training, new samples are generated by drawing $x_0 \sim p_0$ and numerically integrating the learned velocity field forward to $t=1$.

Standard implementations compute $\partial \mathcal{L}_{\text{CFM}} / \partial \theta$ via backpropagation through the velocity network. In Section~\ref{sec:method}, we show how to replace this with Equilibrium Propagation, where the velocity field arises as the energy gradient of a convergent dynamical system and parameter gradients are extracted from equilibrium measurements alone.

\section{Method}\label{sec:method}

\subsection{Flow Matching as Energy Gradient Learning}

In standard flow matching, the velocity field $v_t(x;\theta)$ is parameterised by an unconstrained neural network. We instead define it as the negative energy gradient of a convergent dynamical system:
\begin{equation}\label{eq:v-from-E}
    v(x; \theta) = -\alpha\, \nabla_x E_{\text{int}}(x, h^*(x); \theta),
\end{equation}
where $E_{\text{int}}$ is an internal energy with trainable parameters $\theta$, $h^*(x)$ is the hidden-state equilibrium obtained by holding $x$ fixed and settling the hidden dynamics to $\nabla_h E_{\text{int}} = 0$, and $\alpha$ is an output scale factor.\footnote{When tuned well, $\alpha$ sets the overall velocity magnitude, preventing the optimizer from spending capacity uniformly scaling all weights to fit the magnitude rather than learning the spatial structure of the velocity field.} This defines a velocity field over the full input space, and training shapes the energy gradients so that $v(x;\theta)$ matches the target velocity from flow matching.

During generation, a sample $x \sim p_0$ is evolved by numerically integrating $\dot{x} = v(x; \theta)$. In simulation this is done by re-equilibrating the hidden state $h^*(x)$ at each integration step; on physical hardware, the same effect arises naturally when the hidden dynamics relax on a faster timescale than the visible evolution, corresponding to an adiabatic regime.

Since $\nabla_x E_{\text{int}}$ is time-independent and curl-free by construction, it cannot represent arbitrary time-dependent marginal velocity fields. However, computing an optimal transport coupling between noise and data in mini-batches~\cite{tongImprovingGeneralizingFlowbased2024} --- pairing each noise $x_0$ with the image $x_1$ that minimises total transport distance, rather than pairing randomly --- reduces the overlap between conditional trajectories, making the marginal velocity field more nearly time-independent and curl-free. This approximation has been shown to achieve competitive generation quality in practice~\cite{balcerakEnergyMatchingUnifying2025}. Moreover, a time-independent energy is naturally suited to neuromorphic implementation, as it requires no reconfiguration of hardware parameters across flow times.

The flow matching loss \eqref{eq:cfm-ot} requires $\partial \mathcal{L} / \partial \theta$, i.e.\ gradients of the velocity error with respect to the energy function's parameters. Standard EP computes gradients of a loss on the \textit{settled state}; here, the loss acts on the \textit{energy gradient} at the settled configuration $(x_t, h^*(x_t))$. To extract these gradients via EP, we introduce Gradient Equilibrium Propagation (GradEP).

\subsection{GradEP: Spring-Clamped Equilibrium Propagation}

In standard EP for classification, the input $x$ is clamped (fixed) and only hidden units $h$ evolve. For flow matching, we need the visible units $x$ to evolve as well, so that their equilibrium displacement from the anchor point $x_t$ encodes the learned velocity. We achieve this by replacing the hard clamp on $x$ with a spring potential.

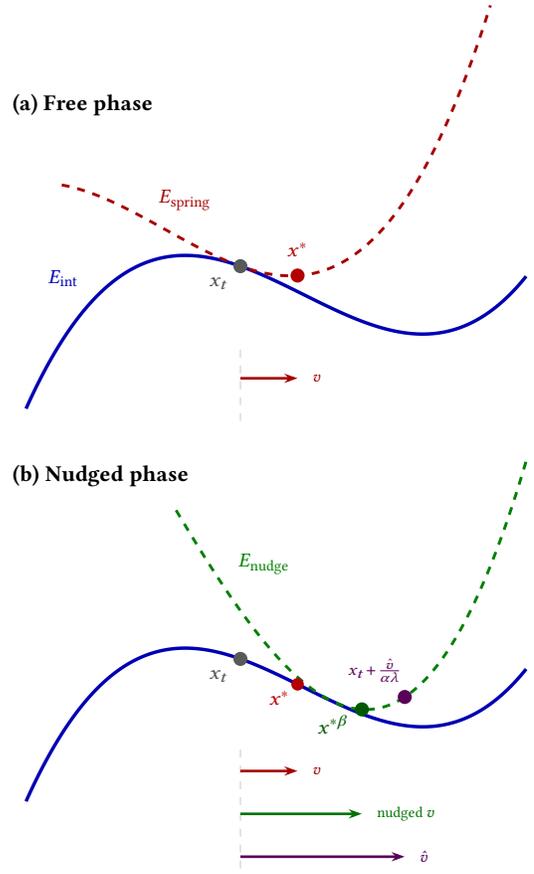
\begin{figure}[t]
    \centering
    \begin{tikzpicture}[
        >=Stealth,
        every node/.style={font=\small},
        scale=0.95, transform shape
    ]

    % ============================================================
    % PANEL (a) — Free phase
    % ============================================================
    \begin{scope}[shift={(0,0)}]

        \def\xt{0.5}
        \def\xstar{1.3}
        \def\Ext{3.67}
        \def\Espring{3.539}

        \node[font=\normalsize\bfseries, anchor=north west] at (-2.8, 6.2) {(a) Free phase};

        % E_int curve
        \draw[thick, blue!70!black, line width=1.3pt] 
            plot[smooth, domain=-2.5:4.5, samples=100] 
            (\x, {0.06*(\x)^3 - 0.25*(\x)^2 - 0.15*(\x) + 3.8});
        \node[blue!70!black, font=\small\itshape, anchor=west] at (-2.3, 3.5) {$E_{\mathrm{int}}$};

        % E_spring curve
        \draw[thick, dashed, red!65!black, line width=1.1pt]
            plot[smooth, domain=-2.0:4.0, samples=100]
            (\x, {0.06*(\x)^3 - 0.25*(\x)^2 - 0.15*(\x) + 3.8 + 0.35*(\x-\xt)*(\x-\xt)});
        \node[red!65!black, font=\small\itshape, anchor=east] at (0.2, 4.6) {$E_{\mathrm{spring}}$};

        % x* marker
        \fill[red!70!black] (\xstar, \Espring) circle (2.8pt);
        \node[red!70!black, font=\small, anchor=south] at ({\xstar}, {\Espring+0.15}) {$x^*$};

        % x_t marker
        \fill[gray!70!black] (\xt, \Ext) circle (2.8pt);
        \node[gray!60!black, font=\small, anchor=north east] at ({\xt-0.05}, {\Ext-0.05}) {$x_t$};

        % Velocity arrow: current v
        \draw[thin, -{Stealth[length=5pt, width=3.5pt]}, red!65!black, line width=1.0pt]
            (\xt, 2.1) -- (\xstar, 2.1);
        \node[red!65!black, font=\scriptsize, anchor=west] at ({\xstar+0.1}, 2.1) 
            {$v$};

        % Thin dashed guideline from x_t down to arrow
        \draw[gray!40, thin, dashed] (\xt, 2.5) -- (\xt, 1.5);

    \end{scope}

    % ============================================================
    % PANEL (b) — Nudged phase
    % ============================================================
    \begin{scope}[shift={(0,-5.5)}]

        \def\xt{0.5}
        \def\xstar{1.3}
        \def\xstarbeta{2.2}
        \def\xanchor{2.8}
        \def\Ext{3.67}
        \def\effcenter{1.8}

        \node[font=\normalsize\bfseries, anchor=north west] at (-2.8, 6.5) {(b) Nudged phase};

        % E_int curve
        \draw[thick, blue!70!black, line width=1.3pt] 
            plot[smooth, domain=-2.5:4.5, samples=100] 
            (\x, {0.06*(\x)^3 - 0.25*(\x)^2 - 0.15*(\x) + 3.8});

        % E_nudge curve
        \draw[thick, dashed, green!50!black, line width=1.1pt]
            plot[smooth, domain=-0.4:4.5, samples=100]
            (\x, {0.06*(\x)^3 - 0.25*(\x)^2 - 0.15*(\x) + 3.8 + 0.4*(\x-\effcenter)*(\x-\effcenter)});
        \node[green!50!black, font=\small\itshape, anchor=east] at (1.3, 5.0) {$E_{\mathrm{nudge}}$};

        % x_t marker
        \fill[gray!70!black] (\xt, \Ext) circle (2.8pt);
        \node[gray!60!black, font=\small, anchor=north east] at ({\xt-0.05}, {\Ext-0.05}) {$x_t$};

        % x* from free phase
        \fill[red!70!black] (\xstar, 3.3145) circle (2.5pt);
        \node[red!70!black, font=\small, anchor=south east] at ({\xstar}, {3.3145-0.4}) {$x^*$};

        % x*beta marker
        \fill[green!35!black] (\xstarbeta, 2.963) circle (2.8pt);
        \node[green!35!black, font=\small, anchor=south] at (\xstarbeta-0.4, {2.963-0.45}) {$x^{*\beta}$};

        % Target anchor
        \fill[violet!70!black] (\xanchor, 3.137) circle (2.8pt);
        \node[violet!70!black, font=\scriptsize, anchor=south west] at ({\xanchor-0.9}, {3.137+0.1}) 
            {$x_t\!+\!\frac{\hat{v}}{\alpha\lambda}$};

        % Thin dashed guideline from x_t
        \draw[gray!40, thin, dashed] (\xt, 2.4) -- (\xt, 0.6);

        % Arrow 1: current velocity
        \draw[thin, -{Stealth[length=5pt, width=3.5pt]}, red!65!black, line width=1.0pt]
            (\xt, 2.1) -- (\xstar, 2.1);
        \node[red!65!black, font=\scriptsize, anchor=west] at ({\xstar+0.1}, 2.1) 
            {$v$};

        % Arrow 2: nudged velocity
        \draw[thin, -{Stealth[length=5pt, width=3.5pt]}, green!45!black, line width=1.0pt]
            (\xt, 1.5) -- (\xstarbeta, 1.5);
        \node[green!45!black, font=\scriptsize, anchor=west] at ({\xstarbeta+0.1}, 1.5) 
            {nudged $v$};

        % Arrow 3: target velocity
        \draw[thin, -{Stealth[length=5pt, width=3.5pt]}, violet!70!black, line width=1.0pt]
            (\xt, 0.9) -- (\xanchor, 0.9);
        \node[violet!70!black, font=\scriptsize, anchor=west] at ({\xanchor+0.1}, 0.9) 
            {$\hat{v}$};

    \end{scope}

    \end{tikzpicture}
    \caption{GradEP mechanism. (a)~Free phase: the spring potential (red, dashed) creates a minimum at $x^*$ near $x_t$; the displacement encodes the current velocity $v$. (b)~Nudged phase: the nudge loss pulls the equilibrium from $x^*$ toward $x^{*\beta}$, closer to the target velocity $\hat{v}$.}
    \Description{Two-panel diagram showing the GradEP mechanism. Panel (a) shows a blue curve representing the internal energy and a red dashed curve representing the spring-clamped energy, with points marking the query point x_t and equilibrium x_star, and an arrow showing the learned velocity proportional to their displacement. Panel (b) shows the nudged phase with a green dashed curve for the nudge energy, showing how the equilibrium shifts from x_star toward x_star_beta, with three horizontal arrows comparing current velocity, nudged velocity, and target velocity.}
    \label{fig:spring-diagram}
\end{figure}

\paragraph{Free phase}
For a given flow-matching training sample, we draw $t \sim \mathcal{U}[0,1]$, data $x_1 \sim q(x_1)$, and noise $x_0 \sim p_0$ paired to $x_1$ via OT coupling, and form the interpolant $x_t = (1-t)x_0 + tx_1$. We then define the spring-clamped energy:
\begin{equation}\label{eq:spring-energy}
    E_{\text{spring}}(x, h; \theta) = E_{\text{int}}(x, h; \theta) + \frac{\lambda}{2}\|x - x_t\|^2,
\end{equation}
where $\lambda$ is the spring stiffness. Starting from $x = x_t$ and $h = 0$, both $x$ and $h$ evolve by gradient descent on $E_{\text{spring}}$, settling to an equilibrium $(x^*, h^*)$. At equilibrium, the visible units satisfy:
\begin{equation}\label{eq:free-equil}
    \nabla_x E_{\text{int}}(x^*, h^*; \theta) + \lambda(x^* - x_t) = 0,
\end{equation}
while the hidden units satisfy $\nabla_h E_{\text{int}}(x^*, h^*; \theta) = 0$. The learned velocity can therefore be read off as:
\begin{equation}\label{eq:velocity-readoff}
    v = \alpha \lambda (x^* - x_t) = -\alpha\,\nabla_x E_{\text{int}}(x^*, h^*; \theta),
\end{equation}
where $\alpha$ is the output scale factor from \eqref{eq:v-from-E}. The spring creates a stable equilibrium for $x$ where EP can operate (without it, $x$ would follow the energy gradient with no fixed point), and the equilibrium displacement $x^* - x_t$ directly encodes the learned velocity (Figure~\ref{fig:spring-diagram}).

Note that \eqref{eq:velocity-readoff} gives the velocity at $x^*$, not $x_t$. Since $x^* = x_t + O(1/\lambda)$ from \eqref{eq:free-equil}, the velocity approximation error is $O(1/\lambda)$, vanishing as $\lambda \to \infty$ (see Appendix~\ref{app:spring-derivation}).

\paragraph{Nudged phase}
At the free equilibrium, the flow matching loss is:
\begin{equation}\label{eq:spring-loss}
    \mathcal{L} = \frac{1}{2}\|v - \hat{v}\|^2 = \frac{(\alpha\lambda)^2}{2}\left\|x^* - \left(x_t + \frac{\hat{v}}{\alpha\lambda}\right)\right\|^2,
\end{equation}
where $\hat{v} = x_1 - x_0$ is the target velocity. Following standard EP, we define a nudge loss as a function of the dynamical variable $x$:
\begin{equation}\label{eq:nudge-loss}
    \ell(x) = \frac{(\alpha\lambda)^2}{2}\left\|x - \left(x_t + \frac{\hat{v}}{\alpha\lambda}\right)\right\|^2,
\end{equation}
which evaluates to $\mathcal{L}$ at the free equilibrium $x = x^*$. Adding $\beta\,\ell$ to the energy gives the nudged energy:
\begin{equation}\label{eq:nudged-energy}
    E_{\text{nudge}} = E_{\text{spring}} + \beta\,\ell(x).
\end{equation}
The nudge term is purely quadratic in $x$ and requires no backpropagation through the energy function, making it readily implementable on neuromorphic hardware. The target anchor $x_t + \hat{v}/(\alpha\lambda)$ is the displacement that would produce the target velocity $\hat{v}$ via \eqref{eq:velocity-readoff}.

Using three-phase symmetric nudging~\cite{laborieuxScalingEquilibriumPropagation2020}, the system is initialised from the free equilibrium $(x^*, h^*)$ and evolves to nudged equilibria $(x^{*\pm\beta}, h^{*\pm\beta})$ under positive and negative $\beta$, and the EP parameter update is:
\begin{equation}\label{eq:ep-spring-update}
    \Delta\theta \propto -\frac{1}{2\beta}\left(\frac{\partial E_{\text{int}}}{\partial \theta}\bigg|_{(x^{*+\beta}, h^{*+\beta})} - \frac{\partial E_{\text{int}}}{\partial \theta}\bigg|_{(x^{*-\beta}, h^{*-\beta})}\right),
\end{equation}
where we use $E_{\text{int}}$ rather than the full energy since neither the spring nor nudge terms depend on $\theta$. Standard EP guarantees that this recovers $\partial \mathcal{L} / \partial \theta$ exactly as $\beta \to 0$. The finite-difference error scales as $O((\beta\alpha^2\lambda)^2)$ rather than the usual $O(\beta^2)$, requiring $\beta\alpha^2\lambda \ll 1$ (see Appendix~\ref{app:spring-derivation}).

The GradEP framework above provides a general mechanism for training energy gradients via EP. This is relevant wherever the learning objective depends on $\nabla_x E$ --- whether interpreted as a velocity field for flow matching, a score function for score matching~\cite{songGenerativeModelingEstimating2019}, or a potential gradient for energy-based generation~\cite{balcerakEnergyMatchingUnifying2025}. As a first demonstration, we apply it to unconditional image generation via flow matching (FlowEqProp), with implementation details described in Section~\ref{sec:experiments}.

\section{Experiments}\label{sec:experiments}

\subsection{Setup}
We first evaluate FlowEqProp on the Optical Recognition of Handwritten Digits dataset~\cite{e.alpaydinPenBasedRecognitionHandwritten1996}, consisting of 1797 $8 \times 8$ greyscale images of digits 0--9, scaled to $[-1, 1]$. We choose this dataset because its low dimensionality reduces the demands on model architecture, allowing us to validate the learning algorithm itself.

We implement $E_{\text{int}}$ as a two-hidden-layer MLP with architecture $64 \to 128 \to 128$ (24,896 trainable parameters) and bilinear inter-layer couplings:
\begin{equation}\label{eq:energy-mlp}
    E_{\text{int}} = \frac{1}{2}\|s\|^2 - \Phi_{\text{coupling}},
\end{equation}
where $s = (x, h_1, h_2)$ comprises visible and hidden units, and the coupling function is:
\begin{equation}\label{eq:coupling}
    \Phi_{\text{coupling}} = b \cdot x + \sigma(h_1) \cdot (W_0 x + b_0) + \sigma(h_2) \cdot (W_1 \sigma(h_1) + b_1),
\end{equation}
with $\sigma$ the SiLU activation applied within the coupling, $b$ a learnable visible bias, and $\{W_l, b_l\}$ the inter-layer weights and biases.

Training uses GradEP with three-phase symmetric nudging and OT coupling over the full dataset. The chosen hyperparameters (Table~\ref{tab:hyperparams}) satisfy $\beta\alpha^2\lambda = 0.075 \ll 1$, as required by the finite-difference accuracy analysis in Appendix~\ref{app:spring-derivation}. Training takes approximately an hour on a single NVIDIA A100 GPU.

\begin{table}[t]
    \centering
    \caption{Hyperparameters for digits generation experiments}
    \label{tab:hyperparams}
    \begin{tabular}{ll}
        \toprule
        \textbf{Parameter} & \textbf{Value} \\
        \midrule
        Architecture & $64 \to 128 \to 128$ \\
        Activation $\sigma$ & SiLU \\
        Weight init & Xavier normal (gain $= 0.5$) \\
        Convergence steps $T$ & 300 (all phases) \\
        Step size $\epsilon$ & 0.1 \\
        Spring stiffness $\lambda$ & 15 \\
        Output scale $\alpha$ & 2 \\
        Nudge strength $\beta$ & $1.25 \times 10^{-3}$ \\
        Optimizer & Adam ($\beta_1\!=\!0.9$, $\beta_2\!=\!0.95$) \\
        Learning rate & $10^{-3}$ \\
        Batch size & 1797 (full dataset) \\
        Training epochs & 2000 \\
        Generation $\mathrm{d}t$ & 0.01 \\
        \bottomrule
    \end{tabular}
\end{table}

Samples are generated by drawing $x_0 \sim \mathcal{N}(0, I)$ and numerically integrating $\dot{x} = v(x; \theta)$ from $t = 0$ to $t = 1$ via Euler steps. At each step, the hidden state $h^*(x)$ is re-equilibrated and the velocity is computed as $v = -\alpha\,\nabla_x E_{\text{int}}$. On neuromorphic hardware, this reduces to letting the physical system relax at each integration step, with the hidden dynamics settling on a faster timescale than the visible evolution.

\subsection{Results}

\begin{figure}[t]
    \centering
    \includegraphics[width=\columnwidth]{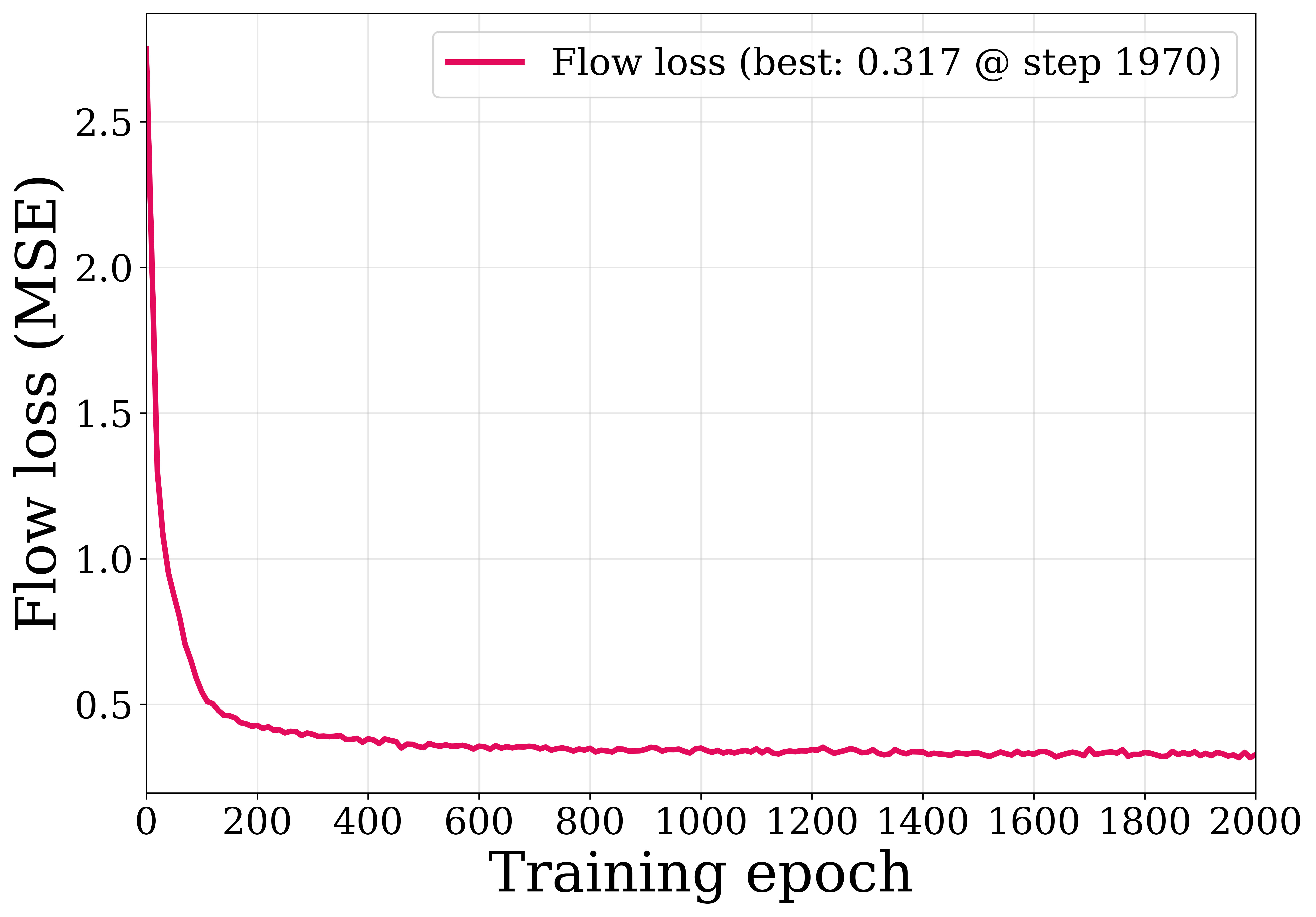}
    \caption{Flow matching loss during training with GradEP. Loss decreases smoothly from 2.75 to 0.32 over 2000 epochs}
    \Description{Line plot showing flow matching loss decreasing smoothly from 2.75 to 0.32 over 2000 training epochs.}
    \label{fig:flow-loss}
\end{figure}

\paragraph{Training dynamics.}
Figure~\ref{fig:flow-loss} shows the flow matching loss over training. The loss decreases smoothly from $2.75$ to $0.32$ over 2000 epochs, demonstrating that GradEP produces stable gradient estimates for flow matching despite relying on finite-difference equilibrium measurements rather than backpropagation.

\begin{figure}[t]
    \centering
    \includegraphics[width=\columnwidth]{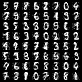}
    \caption{64 generated digit samples from $t = 0$ to $t = 1$. Most samples are clearly identifiable across all ten digit classes}
    \Description{Grid of 64 generated 8 by 8 pixel greyscale digit images showing samples from all ten digit classes, most clearly identifiable.}
    \label{fig:generations}
\end{figure}

\paragraph{Generation quality.}
Figure~\ref{fig:generations} shows 64 samples generated by integrating the learned velocity field from $t = 0$ to $t = 1$. The majority of samples are clearly identifiable as distinct digit classes, with diversity across all ten categories.

\begin{figure}[t]
    \centering
    \includegraphics[width=\columnwidth]{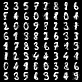}
    \caption{64 generated digit samples from $t = 0$ to $t = 1.2$, showing sharper samples compared to $t = 1.0$ (Figure~\ref{fig:generations})}
    \Description{Grid of 64 generated 8 by 8 pixel greyscale digit images integrated to t equals 1.2, showing sharper edges and higher contrast compared to t equals 1.0 generations.}
    \label{fig:extended}
\end{figure}

\paragraph{Extended generation.}
Because the velocity field is time-independent, integration beyond $t = 1$ simply corresponds to additional gradient descent on the learned energy landscape, rather than querying the model in an untrained regime where sample quality degrades (as for usual time-dependent flow matching). Figure~\ref{fig:extended} shows samples generated to $t = 1.2$. Compared to $t = 1.0$ (Figure~\ref{fig:generations}), the extended samples exhibit sharper edges and more clearly resolved digit structure, as the system settles deeper into energy minima near the data manifold. This is a natural form of adaptive inference-time compute: a fixed model produces higher-quality outputs by expending additional computation, with no retraining required. For neuromorphic hardware, this translates to allowing the physical system to relax longer before reading out the result. Integration well beyond the optimum (e.g.\ $t = 10$) leads to over-sharpening however, suggesting a practical sweet spot near $t \approx 1.2$ for this architecture.

\paragraph{Spring stiffness.}
Spring stiffness $\lambda$ controls the velocity approximation fidelity (Section~\ref{sec:method}). Training is stable across a range of $\lambda$ values, with flow loss improving monotonically: $0.336$ ($\lambda\!=\!3$), $0.329$ ($\lambda\!=\!5$), $0.320$ ($\lambda\!=\!7.5$), $0.314$ ($\lambda\!=\!15$), consistent with the $O(1/\lambda)$ approximation error derived in Appendix~\ref{app:spring-derivation}.

\section{Discussion}\label{sec:discussion}

We have presented FlowEqProp, the first demonstration of Equilibrium Propagation training a flow-based generative model. GradEP provides stable, hardware-plausible gradient estimates for flow matching, requiring only local equilibrium measurements and no backpropagation. The smooth training dynamics and recognisable generations on the digits dataset confirm that the learning algorithm works, establishing a foundation for scaling to more challenging tasks.

The primary bottleneck appears to be architectural rather than algorithmic. The bilinear couplings, combined with the time-inde\-pendent, curl-free formulation, constrain the class of velocity fields that can be represented. However, the Energy Matching framework~\cite{balcerakEnergyMatchingUnifying2025} retains the time-independent, curl-free constraints but achieves state-of-the-art generation quality through more expressive architectures and an additional contrastive divergence objective, suggesting that these constraints need not be prohibitive given sufficient model capacity.

GradEP adds minimal complexity to standard EP: the spring and nudge terms are both quadratic potentials, no more complex than the loss terms used in EP for classification. The $\lambda$ ablation confirms that training is robust across a range of spring stiffness values, which is reassuring for physical implementations where components cannot be tuned to arbitrary precision.

It is notable that, although training only shapes the energy landscape near the interpolation points $x_t$ (since the spring keeps $x^*$ within $O(1/\lambda)$ of $x_t$), generation traverses the full trajectory from noise to data using the bare energy gradient without any spring mechanism, and still produces recognisable samples.\footnote{One might expect that retaining the spring at inference --- computing velocities as $v = \alpha\lambda(x^* - x)$ to match the training dynamics --- would produce better results. In practice, it produces comparable generation quality, suggesting that training successfully shapes the global energy landscape rather than just its local behaviour near $x_t$. This spring-clamped procedure also requires repeated equilibration of both $x$ and $h$ and is less suited to neuromorphic deployment.}

The generation procedure --- in which hidden neurons equilibrate rapidly while visible neurons evolve slowly under the energy gradient --- maps naturally onto neuromorphic hardware with two distinct dynamical timescales. Physical implementations, for instance using GHz-frequency oscillators, could achieve microsecond-scale convergence times $T$~\cite{gowerHowTrainOscillator_Published, main-OIMs-paper}, potentially accelerating generation by orders of magnitude compared to software simulation, where the sequential dynamics cannot be parallelised on conventional hardware, while also consuming a fraction of the energy. The extended generation result (Section~\ref{sec:experiments}) translates directly to this setting: allowing the physical system to relax longer produces sharper outputs, with no retraining required.

Looking forward, the most promising direction is combining GradEP with more expressive EP-compatible architectures such as Energy Transformers~\cite{hooverEnergyTransformer2023} or Modern Hopfield Networks~\cite{ramsauerHopfieldNetworksAll2021}, and incorporating contrastive divergence to shape energy minima at data locations. Scaling to higher-dimensional datasets such as MNIST with convolutional architectures is ongoing work. More broadly, convergent energy-based models have appealing properties for generation but are underexplored in part because their sequential dynamics are slow to simulate on conventional hardware. Neuromorphic implementations, where these dynamics execute natively and efficiently, could make such models practical to develop and deploy at scale.

\begin{acks}
Code is available at \url{https://github.com/alexgower/FlowEqProp}.
\end{acks}

\bibliographystyle{ACM-Reference-Format}
\bibliography{references}  

% \appendices
\appendix
\section{GradEP Derivation}\label{app:spring-derivation}

\paragraph{Free phase equilibrium}
At equilibrium of the spring-clamped energy \eqref{eq:spring-energy}, $\nabla_x E_{\text{spring}} = 0$ gives:
\begin{equation}
    \nabla_x E_{\text{int}}(x^*, h^*; \theta) = -\lambda(x^* - x_t).
\end{equation}
The equilibrium displacement is therefore $x^* - x_t = -\nabla_x E_{\text{int}} / \lambda = O(1/\lambda)$, confirming that $x^*$ approaches $x_t$ as $\lambda \to \infty$.

\paragraph{Velocity approximation error}
The learned velocity at $x^*$ is:
\begin{equation}
    v(x^*) = -\alpha\,\nabla_x E_{\text{int}}(x^*, h^*(x^*); \theta) = \alpha\lambda(x^* - x_t).
\end{equation}
The true flow matching velocity should be evaluated at $x_t$, where the velocity is:
\begin{equation}
    v(x_t) = -\alpha\,\nabla_x E_{\text{int}}(x_t, h^*(x_t); \theta).
\end{equation}

Since our energy function is composed of smooth components (linear/convolutional layers, smooth activations, bilinear couplings), $\nabla_x E_{\text{int}}$ is Lipschitz continuous with some constant $L$, i.e.\ $\|\nabla_x E_{\text{int}}(a) - \nabla_x E_{\text{int}}(b)\| \leq L\|a - b\|$ for all $a, b$. The velocity error is therefore bounded by:
\begin{align}
    \|v(x^*) - v(x_t)\| &= \alpha\|\nabla_x E_{\text{int}}(x^*) - \nabla_x E_{\text{int}}(x_t)\| \nonumber \\
    &\leq \alpha L \|x^* - x_t\| = O(\alpha L / \lambda),
\end{align}
which vanishes as $\lambda \to \infty$.

\paragraph{Nudge loss derivation}
From \eqref{eq:velocity-readoff}, the velocity at the free equilibrium is $v = \alpha\lambda(x^* - x_t)$. The flow matching loss is:
\begin{align}
    \mathcal{L} &= \frac{1}{2}\|v - \hat{v}\|^2 = \frac{1}{2}\|\alpha\lambda(x^* - x_t) - \hat{v}\|^2 \\
    &= \frac{(\alpha\lambda)^2}{2}\left\|x^* - x_t - \frac{\hat{v}}{\alpha\lambda}\right\|^2 \\
    &= \frac{(\alpha\lambda)^2}{2}\left\|x^* - \left(x_t + \frac{\hat{v}}{\alpha\lambda}\right)\right\|^2.
\end{align}
We define the nudge loss as a function of the dynamical variable $x$:
\begin{equation}
    \ell(x) = \frac{(\alpha\lambda)^2}{2}\left\|x - \left(x_t + \frac{\hat{v}}{\alpha\lambda}\right)\right\|^2,
\end{equation}
which evaluates to $\mathcal{L}$ at the free equilibrium $x = x^*$. This is the standard EP construction: the loss is defined as a function of the system state, and adding $\beta\,\ell$ to the energy biases the dynamics toward lower loss. Adding $\beta\,\ell(x)$ to $E_{\text{spring}}$ gives the nudged energy \eqref{eq:nudged-energy}.

\paragraph{Nudged equilibrium derivation}
At the nudged equilibrium, $\nabla_x E_{\text{nudge}} = 0$ gives:
\begin{align}
    &\nabla_x E_{\text{int}}(x^{*\beta}, h^{*\beta}; \theta) + \lambda(x^{*\beta} - x_t) \nonumber \\
    &\quad + \beta(\alpha\lambda)^2\left(x^{*\beta} - x_t - \frac{\hat{v}}{\alpha\lambda}\right) = 0.
\end{align}
Since the spring dominates the energy landscape at large $\lambda$, we can approximate $\nabla_x E_{\text{int}}(x^{*\beta}) \approx \nabla_x E_{\text{int}}(x^*) = -\lambda(x^* - x_t)$; the Hessian correction is $O(\beta)$, negligible compared to the $O(\lambda)$ spring terms. Defining $d = x^* - x_t$ and $d_\beta = x^{*\beta} - x_t$:
\begin{equation}
    -\lambda d + \lambda d_\beta + \beta(\alpha\lambda)^2\left(d_\beta - \frac{\hat{v}}{\alpha\lambda}\right) = 0.
\end{equation}
Collecting terms in $d_\beta$:
\begin{equation}
    \lambda(1 + \beta\alpha^2\lambda)\, d_\beta = \lambda d + \beta\alpha\lambda\hat{v}.
\end{equation}
Rearranging for $d_\beta$:
\begin{equation}
    d_\beta \approx \frac{d + \beta\alpha\hat{v}}{1 + \beta\alpha^2\lambda}.
\end{equation}
The nudge-minus-free displacement is therefore:
\begin{align}\label{eq:nudge-displacement}
    \Delta x &= x^{*\beta} - x^* = d_\beta - d \approx \frac{d + \beta\alpha\hat{v} - d(1 + \beta\alpha^2\lambda)}{1 + \beta\alpha^2\lambda} \nonumber \\
    &= \frac{\beta\alpha(\hat{v} - \alpha\lambda d)}{1 + \beta\alpha^2\lambda} = \frac{\beta\,\alpha\,(\hat{v} - v)}{1 + \beta\alpha^2\lambda},
\end{align}
where we used $v = \alpha\lambda d = \alpha \lambda (x^* - x_t)$ in the last step.

\paragraph{Finite difference accuracy.}
EP's symmetric estimator \eqref{eq:ep-spring-update} is a central finite difference in $\beta$:
\begin{equation}
    \frac{1}{2\beta}\left(\frac{\partial E_{\text{int}}}{\partial \theta}\bigg|_{s^{*+\beta}} - \frac{\partial E_{\text{int}}}{\partial \theta}\bigg|_{s^{*-\beta}}\right) = \frac{d}{d\beta} \frac{\partial E_{\text{int}}}{\partial \theta}(s^{*\beta})\bigg|_{\beta=0} + O(\beta^2).
\end{equation}
The right-hand side equals $\partial\mathcal{L}/\partial\theta$ by EP's theorem. The $O(\beta^2)$ error coefficient depends on the third derivative of $s^{*\beta}$ with respect to $\beta$. From \eqref{eq:nudge-displacement}, with $c := \alpha^2\lambda$:
\begin{equation}
    \frac{d^3}{d\beta^3}\left(\frac{\beta}{1 + c\beta}\right)\bigg|_{\beta=0} = 6c^2,
\end{equation}
so the finite difference error scales as $O(\beta^2 c^2) = O((\beta\alpha^2\lambda)^2)$ rather than $O(\beta^2)$.

Note that the method requires two simultaneous conditions: $\lambda \gg 1$ (so the spring dominates the Hessian of $E_{\text{int}}$, ensuring velocity accuracy at $x_t$) and $\beta\alpha^2\lambda \ll 1$ (for small finite-difference error). These are compatible, requiring $\beta \ll 1/(\alpha^2\lambda)$. In standard EP for classification, the loss $\ell(s) = \frac{1}{2}\|y - \hat{y}\|^2$ has curvature $O(1)$ in the dynamical variable $s$, so $\beta \ll 1$ suffices. In our setting, the nudge loss $\ell(x)$ has curvature $(\alpha\lambda)^2$ in $x$, tightening the requirement to $\beta \ll 1/(\alpha^2\lambda)$.

\end{document}